\documentclass{pasj00}

 \SetRunningHead{H.~Imai \etal}
{Strong $^{13}$CO $J=3\rightarrow 2$ line in \i1634}
 \Received{2011/10/22}
 \Accepted{2012/03/15}
 \def \vlsr{$V_{\mbox{\scriptsize LSR}}$}
 \def \kms{~km~s$^{-1}$}
 \def \etal{~et~al.}
 \def \h2o{H$_{2}$O}
 \def \i1634{IRAS~16342$-$3814}
 
 \def \j32{$J=3\rightarrow 2$}

 \title
{Extremely Strong $^{13}$CO \j32 Line in the ``Water Fountain" \i1634: 
Evidence for the Hot-Bottom Burning}

 \author{Hiroshi  \textsc{Imai}\altaffilmark{1}, 
Sze Ning \textsc{Chong}\altaffilmark{1},  
Jin-Hua  \textsc{He}\altaffilmark{2}, 
Jun-ichi  \textsc{Nakashima}\altaffilmark{3}, 
Chih-Hao \textsc{Hsia}\altaffilmark{3}, \\
Takeshi  \textsc{Sakai}\altaffilmark{4}, 
Shuji  \textsc{Deguchi}\altaffilmark{5}, 
and 
Nico  \textsc{Koning}\altaffilmark{6}
}

 \altaffiltext{1}{Graduate School of Science and Engineering, 
Kagoshima University,  \\
1-21-35 Korimoto, Kagoshima 890-0065}
 \email{hiroimai@sci.kagoshima-u.ac.jp}

 \altaffiltext{2}{Yunnan Astronomical Observatory/ \\
 Key Laboratory for the Structure and Evolution of Celestial Objects,\\
  Chinese Academy of Sciences, Kunming, Yunnan Province 650011, China}

 \altaffiltext{3}{Department of Physics, The University of Hong Kong, Pokfulam Road, 
Hong Kong, China}

 \altaffiltext{4}{Institute of Astronomy, The University of Tokyo,  
2-21-1 Osawa, Mitaka, Tokyo 181-8588} 

 \altaffiltext{5}{Nobeyama Radio Observatory, National Astronomical Observatory, 
Minamimaki, Minamisaku, Nagano 384-1305}

 \altaffiltext{6}{Department of Physics and Astronomy, University of Calgary, Calgary, AB T2N 1N4, Canada}


\begin{document}

 \maketitle

 \begin{abstract}
We observed four ``water fountain" sources in the CO \j32 line emission with the Atacama Submillimeter 
Telescope Experiment (ASTE) 10~m telescope in 2010--2011. The water fountain sources are evolved 
stars that form high-velocity collimated jets traced by  \h2o maser emission. The CO line was detected only 
from \i1634. The present work confirmed that the $^{12}$CO to $^{13}$CO line intensity ratio is $\sim$1.5 at the systemic velocity. We discuss the origins of the very low $^{12}$CO to $^{13}$CO intensity ratio, as possible evidence for the ``hot-bottom burning" in an oxygen-rich star, and the CO intensity variation in \i1634. 
\end{abstract}

 \section{Introduction}

The ``water fountain" sources are a rare group of asymptotic giant branch (AGB) or post-AGB stars that 
show \h2o\ maser emission with a total velocity width larger than that typically seen in 1612~MHz OH 
masers. Previous radio interferometric observations have revealed that the water 
fountains have highly collimated, fast jets traced by \h2o\ maser emission while some of them still have 
circumstellar envelops (CSEs) as seen around AGB and post-AGB stars 
(\cite{ima02,cla09,wal09,day10}, see also a review of \cite{ima07}). It has been 
suggested that the dynamical ages of the jets are shorter than 100~years \citep{ima07}.  
Thus it is expected that the water fountains should shed light on the mechanism of jet launching 
found in planetary nebulae (PNe) and on that of the formation of asymmetric 
PNe (e.g. \cite{sah98}). However, because the volume of the maser emission regions is 
quite limited, observations of thermal emission such as CO and dust continuum are essential 
for understanding the whole spatio-kinematical structures of the water fountains. 

\citet{he08} and \citet{ima09} reported the first example of CO emission towards a water fountain 
source \i1634 (hereafter abbreviated as I16342), whose CO emission lines 
($J=2\rightarrow 1$ and \j32 respectively) were detected in single-dish observations. The 
failure of CO emission detection towards other water fountains \citep{ima09} may be attributed 
to the following factors; they are located close to 
the Galactic plane with heavy contamination from the interstellar CO emission, they are too distant 
($D \gtrsim$2~kpc), or the observed coordinates had large offsets from the true coordinates of 
the sources. Nevertheless, we can learn interesting properties of the CO emission from 
I16342. \citet{he08} found a very low $^{12}$CO to $^{13}$CO line intensity ratio ($\sim$1.7).  
The $^{12}$C/$^{13}$C abundance ratio is not only much lower than those towards interstellar 
clouds ($\sim$70, e.g., \cite{mil05}) but also lower than other AGB/post-AGB stars (e.g., \cite{sch00}). 
\citet{ima09} also detected high-velocity wings in the $^{12}$CO \j32 spectrum, 
whose total velocity range is comparable to that of the \h2o\ maser spectrum\footnote
{In this paper, the CO spectrum published in \citet{ima09} is revised. See Sect. \ref{sec:results}.}. 

In this paper, we report the results of additional observations of the CO \j32 lines towards I16342 
with the Atacama Submillimeter Telescope Experiment (ASTE) 10~m telescope. They were 
conducted to reconfirm the low $^{12}$CO to $^{13}$CO line intensity ratio reported by \citet{he08} 
but for the \j32 lines. 

 \section{Observations and data reduction}

The new ASTE observations of the CO \j32 emission were conducted during LST 
16:00--22:00/20:00--22:00 on 2010 August 16/17 and LST 13:00--21:00 on 2011 June 10. 
In 2010, the $^{12}$CO \j32 emission at 345.79599~GHz was observed towards four water 
fountain sources listed in Table \ref{tab:water-fountains}. In 2011, $^{13}$CO \j32 emission 
at 330.587957~GHz was observed towards only I16342 so as to achieve a spectral noise level 
roughly equal to that of $^{12}$CO \footnote{
We tried to observe the $^{13}$CO line on 2010 August 16, but the obtained data should be 
dropped out because later we confirmed that we observed it in a wrong frequency setup.}.
The FWHM beam sizes of the ASTE telescope are 22\arcsec\ and 23\arcsec\ at the observed 
frequencies of 345~GHz and 330~GHz, respectively. 
The system temperature was between 210 and 420~K (single side band). 
The received signals were down-converted in frequency and transferred into three (in 2010) and 
four (in 2011) base band channels (BBCs), each of which has a band width of 512~MHz, corresponding 
to a velocity width of 445 \kms \ at 345~GHz. In 2010, the center frequencies of the BBCs were split by 
100~MHz  to check the intrinsic CO emission, which should be detected in different spectral channels 
but at the same velocity ranges in the all BBCs\footnotemark[2]. 
In 2011, the center frequencies of the BBCs were set 
to the rest frequencies of $^{12}$CO, $^{13}$CO,  C$^{18}$O (\j32), HCN, H$^{13}$CN, 
HC$^{15}$N ($J=4\rightarrow 3$), SO$_{2}$ (at 331.580~GHz)), CS, HCO$^{+}$, H$^{13}$CO$+$ 
($J=4\rightarrow 3$), and SiC$_{2}$ (at 357.473~GHz). However, except CO, no detection was 
confirmed in the 3-$\sigma$ upper limit to 12--16~mK. This paper reports the results of the 
$^{12}$CO and  $^{13}$CO observations. We used the MAC spectrometer to 
obtain a spectrum with 1024 spectral channels, corresponding to a frequency and velocity spacings of 
500~kHz and 0.45 \kms, respectively. 

An antenna pointing check was made before every change from one target source to another. 
We used the CO emission towards the AGB stars II~Lup and W~Aql. 
The pointing offset derived in these measurements was always within $  \sim$2\arcsec, indicating 
that the pointing stability may be equal to this typical offset value in most of the observations. 
For antenna temperature calibration, 
$^{12}$CO and $^{13}$CO lines towards W28 were observed. Reduction of the spectral data was 
made using the NewStar package developed at the Nobeyama Radio Observatory. 

Nine consecutive spectral channels were re-binned to a velocity resolution of 4.1\kms\ to improve the  
signal-to-noise ratio. Comparing the antenna temperatures of the W28 CO lines with those obtained with 
the Caltech Submillimeter Observatory (CSO) 10~m telescope \citep{wan94} and assuming an antenna 
aperture efficiency of ASTE ($  \eta_{ \mbox{ \scriptsize{MB}}}=$0.59, \cite{ima09}),  we set conversion 
factors of antenna temperature scales in the BBCs. The original temperature scales were multiplied by these 
factors with values within a range of 1.0--1.3 to obtain the corrected scales. The emission-free baseline 
has a linear gradient for the integrated spectrum of \i1634, and it was removed to obtain the final 
spectrum. On the other hand, gentle standing waves can be seen in the spectra of other 
sources. A higher order polynomial baseline was therefore removed. The baseline removal did not 
affect the identification of line features in the spectra because the wavelength of the standing waves is 
usually much broader than a typical line width ($\Delta V\leq$50\kms). 

Table  \ref{tab:water-fountains} gives parameters of the water fountain sources and the observations. 
The observations were made in the antenna-position switching mode; the number of points observed 
on and around the target source ($>$1 for the cross-scan mode) is given in Column 10. Because all 
the sources except \i1634 are located very close to the Galactic plane with intense background 
molecular emission, five-points or nine-points cross scans were adopted, in which the observed 
points were separated by 10\arcsec --20\arcsec (see sect.  \ref{sec:results}). Column 11 in table 
\ref{tab:water-fountains} gives root-mean-square noise levels of the spectra.  

 \section{Results}
 \label{sec:results}

We searched for $^{12}$CO \j32 emission towards IRAS~18043$-$2116 (OH~009.1$-$0.4),  
IRAS~18139$-$1816 (OH~12.8$-$0.9), and  IRAS~19190$+$1102 with the ASTE telescope. 
However, after removing interstellar contamination by subtracting the off-point spectra in our 
cross-scans, no intrinsic CO emission could be recognized around the expected systemic velocities. 
The upper limit of CO \j32 emission is given by 3 times the root-mean-square noise level shown in 
Table \ref{tab:water-fountains}.

Figure \ref{fig:I16342}a shows the spectra of the $^{12}$CO and $^{13}$CO \j32 lines toward \i1634. 
To obtain these spectra with improved sensitivity in 2010, two or three MAC spectra covering different 
velocity ranges were synthesized\footnote
{In this step, we realized that the spectrum shown in \citet{ima09} was incorrectly synthesized from 
the spectra obtained from three BBCs. For two out of the three spectra covering different velocity 
ranges, the frequency scale was incorrectly converted to the velocity scale, leading to a 
velocity-averaged spectrum exhibiting a velocity width and a peak antenna temperature larger and 
less than the true ones, respectively. In this paper, the correctly synthesized spectrum is displayed.}. 
The spectrum obtained in 2011 is synthesized from only the data at the source position, and shown in 
Figure \ref{fig:I16342}a. 
The $^{12}$CO spectra obtained in the three observations seem to be composed of two components: 
a wide wing (\vlsr$\lesssim$15\kms\ and \vlsr$\gtrsim$75\kms) and a central sharp peak 
(15$\lesssim$\vlsr$\lesssim$75 [\kms]). They look roughly symmetric around the systemic velocity 
(\vlsr $\simeq$45\kms) but with some deviation as mentioned later. Although cross-scan observations 
were performed, the CO emission was detected only at the position of I16342 (Figure \ref{fig:I16342}b). 
Although the emission seems to be marginally detected at the points 12\arcsec\ away from the source 
position, it does not mean the existence of extended components because even a perfect point 
source should be detected at these locations.  The peak antenna temperature seems to exhibit  
temporal variation, whose origin is discussed in Sect. \ref{sec:variation}. 
The narrower and weaker $^{13}$CO emission is also detected at the target position. 
The peak velocities of the $^{13}$CO 
spectrum seems to be blue-shifted from that of $^{12}$CO by $\sim$8\kms. Although it is difficult to 
judge whether this shift is real, the comparison of the $^{12}$CO spectrum with the $^{13}$CO one 
supports the shift and the asymmetry of the former relative to the latter. They may indicate 
the optically-thick property of the $^{12}$CO line, which is discussed in Sect. \ref{sec:ratio}. 

Table \ref{tab:i1634} gives parameters of Gaussian fitting to the observed spectra in the cases 
assuming one and two Gaussian components. Interestingly, the $^{12}$CO to $^{13}$CO line 
intensity ratio is very low ($\sim$1.5) at the near systemic velocity. The derived intensity ratio 
is roughly consistent with that for the CO $J=2\rightarrow 1$ lines \citep{he08}. 
The $^{13}$CO line has a velocity width 
a little larger than that of the narrow peak component of the $^{12}$CO spectrum. However, it is 
difficult to recognize the existence of high velocity wings in the $^{13}$CO spectrum. 

 \section{Discussion}

\subsection{Origins of the high-/low-velocity components in the CO spectra}
\label{sec:spectra}

The $^{12}$CO line profiles, as shown Figure \ref{fig:I16342}, exhibit both of a narrow component 
($V_{\rm exp}\lesssim$40\kms) and very wide wings ($V_{\rm exp}\gtrsim$100\kms). The one-Gaussian 
model has large deviation over the 3-$\sigma$ noise level from the observed spectrum. In the 
case of an optically thick 
spherically-expanding flow as seen in CSEs of AGB stars, an observed CO emission shows a 
parabolic spectral profile (c.f., \cite{kem03}). In contrast, the $^{12}$CO line profile towards I16342 
resembles a Gaussian rather than parabolic shape. Taking into account the models for different 
opacity cases (e.g., \cite{de10}), it suggests that at least the 
CO emission associated with the high velocity jet has a Gaussian profile and should be unresolved and 
optically thin. In fact, the jet and the equatorial flow should be 
smaller than the ASTE beam \citep{dij03, sah05, ver09}. The existence of a central dark lane has 
also been confirmed in optical images \citep{sah99,sah05}. Therefore it is reasonable to assume 
that the observed CO emission in I16342 comes from both of a fast bipolar flow and a slowly 
expanding torus/CSE around I16342. 

On the other hand, the $^{13}$CO line profiles, as shown Figure \ref{fig:I16342}, exhibit only a 
narrower component. If the both of the $^{12}$CO and $^{13}$CO lines have the same line 
profile, high-velocity wings of the latter line should also be detected over the 3-$\sigma$ noise level. 
Actually the absence or weakening of such wing components in the latter line is recognized. 
If supposing a common $^{12}$CO/$^{13}$CO abundance ratio in the whole observed region, 
some opacity effects should be considered to explain the difference in the 
$^{12}$CO and $^{13}$CO profiles.  

Using the morpho-kinematic software {\it SHAPE} \citep{ste06,ste11}, we have reproduced both the 
physical and kinematic structure of I16342 with a single model.  To constrain the geometrical and 
physical parameters, we have made use of the near-infrared image of \citet{sah05}. They suggest 
the image represents light from a central star scattered by dust in dense shells surrounding 
a tenuous bipolar cavity. Our model therefore consists of a bipolar cavity extending 1\arcsec\ 
on either side of the central star embedded within a spherical halo of radius 
$r_{\rm halo}=$1.5\arcsec\ halo.  The halo contains both CO and dust and has a density profile 
$\rho{\rm (halo)}=\rho_{\rm 0,halo}(r/r_{\rm 0,halo})^{-2}$ ($0\leq r \leq r_{\rm 0,halo}$). 
The surrounding cavity is a thin, dense shell of CO and dust; presumably swept up by the high 
velocity jet.  The cavity is inclined at $i=$30\arcdeg to the observer with a PA of 67\arcdeg. The 
geometry of the model is displayed in Figure \ref{fig:SHAPE-geometry}. 

Using radiative transfer in {\it SHAPE}, to simulate the emission, absorption, and scattering for  
dust and the first two kinds of molecular particles with different opacities in a given temperature 
(see \cite{ste11} for the technical details), this model produces the image displayed in Figure 
\ref{fig:SHAPE-spectrum}a.  The model is able to reproduce the general appearance of 
I16342 seen in the image from \citet{sah05} (Figure \ref{fig:SHAPE-spectrum}b) in scattered 
light including the extent of both the lobes and dark equatorial waste. 
To constrain the kinematics of I16342, we use the results from our current study. 
The CO emission in our model originates in the dense shell surrounding the cavity, 
and from the extended spherical halo. The velocity within the shell has the form
$V_{\rm exp}{\rm (jet)}=$300\kms[$r/$1\arcsec]. Since the cavity extends out to 1\arcsec,  
the maximum speed of the jet is 300\kms\ at the tip (c.f. \cite{cla09}). The spherical halo is 
given a constant radial velocity of 15\kms. With these kinematic parameters and the 
aforementioned physical model, {\it SHAPE} is able to reproduce the spectra for $^{12}$CO and 
$^{13}$CO in Figure  \ref{fig:SHAPE-spectrum}c and d, respectively. We find that because of 
the absence of high velocity wings in the latter spectrum, $^{13}$CO, a tracer of higher 
density gas, is not bright in the high-density shell, but only in the halo. $^{12}$CO, 
on the other hand, is bright in both the shell and the halo. This suggests that a higher intensity 
ratio of $^{12}$CO to $^{13}$CO may exist in the shell where high temperatures ($>$600K, 
see Sect. \ref{sec:ratio}) are expected and \h2o\ and OH maser actions are excited.  

\subsection{Extremely low $^{12}$CO to $^{13}$CO intensity ratio}
\label{sec:ratio}

We confirmed a very low $^{12}$CO/$^{13}$CO line intensity ratio ($\sim$1.5) for the \j32 and 
$J=2\rightarrow 1$ \citep{he08} transitions towards I16342. At first, one may suppose that this is 
attributed to an opacity effect of these lines. {To examine such a possibility, in other words 
whether even a typical $^{12}$CO/$^{13}$CO abundance ratio as seen in interstellar clouds 
($\sim$70, e.g., \cite{mil05}) can explain such a low line intensity ratio, we performed another 
radiative-transfer simulations that repeatedly calculated the line opacities and the line intensity ratio. 
Note that {\it SHAPE} does not constrain CO line opacities properly because a line opacity in it 
is manually tuned to reproduce the observed brightness distribution of the line/continuum emission 
from the morpho-kinematical model. Thus we resort to another radiative transfer code {\it RADEX}
\footnote[1]{http://www.strw.leidenuniv.nl/\~{ }moldata/radex.html}
\citep{van07}, which makes use the molecular line database {\it LAMDA} \citep{sch05} and treats 
the radiative transfer and CO rotational level excitation in a self-consistent way. 
After many trials with changing input physical parameters such as 
volume density ($10^{2}{\rm ~cm}^{-3}<n_{{\rm H}_{2}}\leq 10^{6}{\rm ~cm}^{-3}$), 
column density ($10^{12}{\rm ~cm}^{-2}<N_{\rm CO}\leq10^{18}{\rm ~cm}^{-2}$), and 
kinematic temperature (10~K$<T_{\rm k}\leq$1000~K), the results of the calculations are 
summarized as follows. Here one assumes a velocity width of $\Delta V\simeq$10\kms\ and 
$N_{\rm CO}/N_{{\rm H}_{2}}\simeq 10^{-4}$. In the case of I16342, the spectral energy 
distribution (SED) indicates a blackbody temperature of the CSE around $T\simeq$130~K, 
which was derived from fitting of a single temperature SED to the data of IRAS\footnote[2]
{Infrared Astronomical Satellite, http://irsa.ipac.caltech.edu/data/download/IRAS/iras\_psc.tbl}, 
MSX6C\footnote[3]
{Midcourse Space Experiment, http://irsa.ipac.caltech.edu/data/MSX/docs/MSX\_psc\_es.pdf}, and 
AKARI\footnote[4]{http://darts.isas.jaxa.jp/astro/akari/cas.html} Point Source Catalogs. 
Taking into account the size of the mid-infrared emission \citep{ver09} and the present ASTE 
observations, which gives an upper limit to the source size comparable to the ASTE beam size 
($<$20\arcsec, the CSE/jet may have a size in the range of 6,000--10,000~AU. 

First, in the case of the volume density of hydrogen molecules close to the critical value, 
$n_{{\rm H}_{2}}\simeq 2\times 10^{5}{\rm ~cm}^{-3}$, the CO \j32\ lines become optically thick 
($\tau\simeq 1$) in a cloud with 
$N_{\rm CO}\gtrsim 2\times 10^{17}{\rm cm}^{-2}$ or $T_{\rm k}\gtrsim$550~K. 
The former threshold value corresponds to a cloud size of $\sim$1000~AU and a mass loss rate 
of $\sim 3\times 10^{-5}M_{\odot}{\rm yr}^{-1}$. 
For the CSE of I16342 with the larger size and the higher temperature as mentioned above, 
the $^{13}$CO opacity is estimated to be comparable to or lower than unity. On the other hand, 
the $^{12}$CO emission has a lower critical density and may have a larger distribution 
including the jet, therefore, the emitting region of the $^{12}$CO line may have a lower temperature 
and the $^{12}$CO line may be optically thick. This is consistent with the assumption of such 
opacity difference from our {\it SHAPE} simulation (see Sect. \ref{sec:spectra}). 
Second, the $^{12}$CO/$^{13}$CO line 
intensity ratio may have the observed small value {\it only} in the case 
where $T_{\rm }<$20~K or $T_{\rm }>$550~K. Note that such low and high temperatures may 
appear, respectively, in the outer boundary of a CSE with a volume density lower than expected 
to excite the CO \j32 emission and at the innermost part of a CSE (e.g., within 100 stellar radii, 
\cite{coo85}). Although tiny clumps to excite \h2o\  and OH maser meet such a physical condition, 
their volume fraction in the CSE should be extremely limited. Regardless, these temperature 
regions should be uncommon in the CSE. Third, in the physical conditions expected for 
the CSE of I16342, even if the $^{12}$CO line is optically thick, a CO column density of 
$N_{\rm CO}>10^{19}{\rm ~cm}^{-2}$ is necessary to explain the observed intensity ratio if 
the CSE has a $^{12}$CO/$^{13}$CO abundance ratio as seen in interstellar 
clouds ($\sim$70). For a moderate value ($N_{\rm CO}\lesssim 10^{18}{\rm ~cm}^{-2}$), 
a much lower abundance ratio, 4 or lower, is required. These results support, the extremely 
low $^{12}$CO/$^{13}$CO abundance ratio in the CSE. 

Applying a frequency correction described in Equation 15 of \citet{de10}, we derive a 
$^{12}$C/$^{13}$C isotopic ratio of $\sim$1.3.  Such an extremely low isotopic ratio has also 
been confirmed in $J$-type carbon-rich stars \citep{abi00, olo99}. The $J$-type star has a lower 
stellar mass ($M_{\ast}\simeq$2--3$M_{\odot}$) and the ``cold-bottom processing" is expected, 
in which production of $^{13}$C is enhanced, but not O (e.g., \cite{abi00}). On the other hand, 
an oxygen-rich star has a higher mass ($4M_{\odot}\lesssim M_{\ast}\lesssim 7M_{\odot}$) and 
the ``hot-bottom burning" (HBB) is expected, in which production of both $^{13}$C and O is 
enhanced so that $^{12}$C is converted to these nucleons (see e.g. a review of \cite{her05, de10}). 
Therefore, the enhancement of $^{13}$CO relative to $^{12}$CO observed in O-rich stars should 
provide direct evidence for HBB in the AGB nucleosynthesis. The present results towards 
I16342, that clearly has O-rich chemistry harboring \h2o\ and OH masers, may provide one of 
the most clear examples of the HBB signature. 
Some pre-planetary nebulae with O-rich chemistry show similar cases and their progenitors 
are also suggested to be massive post-AGB stars (see also e.g., \cite{nak04,nak06,din09}). 
Based on the stellar luminosity and the orbit in the Galaxy, it has been demonstrated that the 
water fountains should harbor intermediate-mass 
O-rich evolved stars such as OH/IR stars and their posterity (e.g., \cite{ima02, ima07, ima07b}). 
The possibility of HBB is consistent with this view. Direct determination of the original mass of 
the central star of I16342 is a future issue in theoretical and observational works.

\subsection{Temporal variation of the CO profile}
\label{sec:variation}

In general it is difficult to precisely compare the peak (main beam) antenna temperatures, 
$T_{\rm MB}$, found in different observation epochs. In the present observations, the peak value 
of $T_{\rm MB}$ is apparently enhanced by a factor of $\sim$1.6 from the first to second epoch 
spectrum. Note that these $T_{\rm MB}$ scales had been corrected by using the different flux 
calibrators (IRC$+$10216 in 2008 and W28 in 2010 and 2011, respectively) while the scaling 
correction factors were smaller than 1.3 in the spectrum calibration of the whole observations. 
It has been expected that the flux calibrators provide $T_{\rm MB}$ scales with uncertainty of 
$\sim$20\% (e.g. \cite{wan94}). It has not yet been reported that these calibrator spectra are 
variable. Therefore the intrinsic temporal variation of the I16342 CO spectrum cannot be ruled out. 

However, we note that when the $T_{\rm MB}$ values are set to equal among the observations, 
the rescaled spectral profiles resemble each other in the whole velocity range within rms noise 
levels. Figure \ref{fig:I16342}c displays the $^{12}$CO spectra obtained in 2008 June \citep{ima09}, 
2010 August, and 2011 June. For comparison of the whole spectra, the spectra in 2008 and 2011 
are rescaled by factors of 1.6 and 1.3, respectively. 
As discussed in Sect. \ref{sec:spectra}, the CO emission may originate from both the fast jet 
and the slowly expanding CSE/torus, which should be physically independent. Taking into account 
a possible time scale of flux variation due to discontinuous stellar mass loss or episodic 
events for mass eruption as well as large physical sizes of the CO sources ($>$1000~AU), 
it is difficult to explain the $T_{\rm MB}$ variation by intrinsic temporary variation. Alternatively, 
different antenna pointing offsets in the observations are expected to explain the similarity of the 
CO spectral profiles and the different antenna temperature scales. If the CO region is an 
ideal point source, such a large variation cannot be expected because the pointing offsets 
should be much smaller than the beam size ($\lesssim$5\arcsec). A simple Monte-Carlo simulation 
which decreased the antenna temperature 
scale suggested that, for the ASTE's beam (22\arcsec), the source size should be larger than 
5\arcsec (when assuming a point-symmetric, Gaussian brightness distribution) to reproduce the 
observed scale decrease. Figure \ref{fig:I16342}b shows the spectra obtained from five-point scans 
conducted in 2011. The CO emission was detected only on the on-source point and the peak values 
of $T_{\rm MB}$ at the off-point positions are $\sim$40\% of that at the on-source point or lower. 
This indicates that the angular size of the CO emission, at least the low velocity component, should 
be smaller than $\sim$10\arcsec. The true angular scale of the observed CO emission is easily 
measured by future interferometric observations. 

\subsection{Mass loss rate of \i1634}
\label{sec:m-dot}
In this paper, we reestimate a mass loss rate of the I16342 CO \j32 outflow from the revised 
$^{12}$CO profile. With assumption of unresolved, optically thick CO emission, as mentioned in 
Sect. \ref{sec:ratio}, we can derive a mass loss rate in units of solar masses per year using the 
formula as follows (\cite{ram08}, see also \cite{kna85, olo93, gro99, de10}), 

\begin{equation}
\dot{M}=s_{J}\left(I_{\rm CO}B^{2}D^{2}\right)^{a_{J}}V_{\rm exp}^{b_{J}}f^{-c_{J}}_{\rm CO}. 
\label{eq:mass-loss}
\end{equation}

 \noindent
Here $I_{\rm CO}$ is the velocity-integrated antenna temperature of the CO emission in K\kms, 
$V_{\rm  exp}$ the expansion velocity of the CO emission in\kms, $D$ the source distance in kpc, $B$ 
the beam size of the telescope in arcsec, $f_{\rm CO}$ the abundance of CO molecules relative to 
H$_{2}$, and $s_{J}\simeq 3.8\times10^{-11}$ a correction factor for $J\rightarrow J-1$ transition, 
and $a_{J}\simeq$ 0.91, $b_{J}\simeq$ 0.39, $c_{J}\simeq$ 0.45 the coefficients derived for \j32, respectively. 
For I16342, $D=$2~kpc and  is adopted. For ASTE, $B=$22\arcsec\ is adopted. 
For an O-rich circumstellar envelope harboring  \h2o \ and OH maser emission, $f_{\mbox{\scriptsize{CO}}}
 \sim 10^{-4}$ is adopted. Using the Gaussian profile parameters 
listed in Table \ref{tab:i1634} (here $V_{\rm exp}\simeq \Delta V_{\rm HWHM}$), the contributions to 
the mass loss rates from the broad-wing and the central-peak components are calculated to be 
$\dot{M}\simeq 4.8 \times 10^{-5}M_{\odot}{\rm yr}^{-1}$ and 
$\dot{M}\simeq 3.8\times 10^{-6}M_{\odot}{\rm yr}^{-1}$, respectively. 
The former value is much smaller than estimated with mid-IR emission 
($\dot{M}_{ \mbox{\scriptsize{gas}}}  \approx  10^{-3} \:M_{\odot}$~yr$^{-1}$, \cite{dij03}) but 
indicates a major contribution to the total mass loss rate of I16342. However, 
the CO emission contribution from the former component should be more carefully examined by high resolution mapping. 

 \section{Conclusions}

Through a series of our ASTE observations, we detected the intrinsic CO emission from one source, 
I16342, out of 13 water fountain sources. We find a very low $^{12}$CO to $^{13}$CO line intensity 
ratio ($\sim$1.5). This may indicate an intrinsic property of the CO emission towards the water fountain 
and indicates the presence of ``hot-bottom burning" in the stellar nucleosynthesis. 
Based on arguments from e.g. \citet{ima07}, the present results support that I16342 should be an 
oxygen-rich star with a 
mass of $4M_{\odot}\lesssim M_{\ast}\lesssim 7M_{\odot}$. I16342 may have a mass loss rate of 
$ \dot{M}_{ \mbox{\scriptsize{gas}}}  >5\times10^{-5} \:M_{ \odot}$~yr$^{-1}$. The detection of the 
high-velocity wings in the $^{12}$CO spectrum implies a possibility that the bipolar high-velocity jet 
plays a major role of final stellar mass loss. High angular resolution and sensitivity CO mapping 
observations as those conduced with the Atacama Large Millimeter-submillimeter Array (ALMA) 
should dramatically increase the number of CO sources towards the water fountains  within a few 
kilo parsecs with moderate mass loss rates as AGB stars ($>10^{-6}M_{\odot}{\rm yr}^{-1}$) and 
spatially resolve the high-velocity wing components in the CO spectra. 

 \bigskip

We deeply appreciate members of the ASTE team for their careful observations, preparation and kind 
operation support. The ASTE project is driven by Nobeyama Radio Observatory (NRO), a branch of 
National Astronomical Observatory of Japan (NAOJ), in collaboration with University of Chile, and Japanese 
institutes including University of Tokyo, Nagoya University, Osaka Prefecture University, Ibaraki University, 
and Hokkaido University. Observations with ASTE were in part carried out remotely from Japan by using 
NTT's GEMnet2 and its partner R\&E (Research and Education) networks, which are based on AccessNova 
collaboration of University of Chile, NTT Laboratories, and NAOJ. We thank T. Minamidani and H. Izumiura 
for introducing the radiative transfer calculation code (RADEX) and for giving fruitful 
comments on our work, respectively. 
HI and SD have been financially supported by Grant-in-Aid for Scientific Research from Japan Society 
for Promotion Science (20540234). JN is supported by a grant from the Research Grants Council of 
Hong Kong (project code: HKU 704209P; HKU 704710P; HKU 704411P) and by the financial support 
from the Small Project Funding of HKU (project code: 201007176004).

 \clearpage


 \begin{table*}[h]
 \caption{Parameters of the water fountain sources and the ASTE observations in $^{12}$CO emission.}
 \label{tab:water-fountains}
 \begin{tabular}{lrl@{}c@{}rrrcrl@{ }rcr}
 \\  \hline  \hline
IRAS name & R.A. (J2000.0) & Year of & $l$\footnotemark[1] & $V_{\mbox{\scriptsize sys}}$\footnotemark[2] 
& $  \Delta V_{\mbox{\scriptsize los}}$\footnotemark[3] 
& $D$\footnotemark[4] & $t_{\mbox{\scriptsize jet}}$\footnotemark[5] & 
& Dur\footnotemark[7] & On-\footnotemark[8] & rms  \\
Other name & Decl. (J2000) & observation & (\arcsec) &  \multicolumn{2}{c}{(km~s$^{-1}$)} 
& (kpc) & (year) & Ref.\footnotemark[6] & (hr) & point & (mK)  \\  \hline

16342$-$3814 & 16$^{h}$37$^{m}$39$^{s}$\hspace{-2pt}.91 
& 2010 & 2.4 & 50 & 240 & 2.0 & 100 & 2, 5  & 0.5 & 1 & 5 \\
OH~344.1$+$5.8 & $-$38\arcdeg 20\arcmin 17\arcsec\hspace{-2pt} .3 & 2011 & & & & & & & 2.4 &  5 & 7 \\ \hline

18043$-$2116 & 18$^{h}$07$^{m}$20$^{s}$\hspace{-2pt}.85 
& 2010 & 0.3 & 87 & 400 & 6.4 & 60 & 3, 7 & 1.25 & 5 & 6 \\
OH~009.1$-$0.4 &  $-$21\arcdeg 16\arcmin 12\arcsec\hspace{-2pt} .0 \\ \hline

18139$-$1816 &  18$^{h}$16$^{m}$49$^{s}$\hspace{-2pt}.23  
& 2010 & 0.12 & 56 & 50 & $\sim$8 & 90 & 1 & 2.25 & 5 & 6 \\
OH~12.8$-$0.9 &  $-$18\arcdeg 15\arcmin 01\arcsec\hspace{-2pt} .8\\ \hline

19190$+$1102 &  19$^{h}$21$^{m}$25$^{s}$\hspace{-2pt}.09
& 2010 & 0.28 & 28 & 130 & 8.6 & 60 & 4, 6 & 0.58 & 5 &  9 \\  
& $+$11\arcdeg 08\arcmin 41\arcsec\hspace{-2pt} .0  & & & & & & & & 1.08 & 9 & 21 \\ \hline
 \end{tabular}

 \noindent
 \footnotemark[1]Total angular length of the jet system.  \\
 \footnotemark[2]Systemic velocity of the jet system.  \\
 \footnotemark[3]Full range of the line-of-sight velocities of \h2o maser emission.  \\
 \footnotemark[4]Distance to the source.  \\
 \footnotemark[5]Dynamical age of the jet 
($ \approx l/ \Delta V_{\mbox{\scriptsize los}}$).  \\
 \footnotemark[6]References of the jet parameters. 
1: \citet{bob07}; 2: \citet{cla09}; 3:  \citet{dea07}; 4 \citet{day10}; 5:  \citet{ima09}; 
6: \citet{sua08}; 7: \citet{wal09}. \\
 \footnotemark[7]Duration of the total observation time with ASTE.  \\
 \footnotemark[8]Number of points observed on and around the target, except an off-point.  \\

\caption{Parameters of Gaussian fitting to $^{12}$CO and $^{13}$CO emission toward \i1634.}
\label{tab:i1634}
\begin{tabular}{lcccc@{\hspace*{3mm}}cccccc}\hline\hline
 & \multicolumn{3}{c}{Single Gaussian fitting} &  \multicolumn{6}{c}{Two Gaussian fitting} \\
 & \multicolumn{3}{c}{\hrulefill} &  \multicolumn{6}{c}{\hrulefill}  \\
 & & & & \multicolumn{3}{c}{Broad wing} &  \multicolumn{3}{c}{Narrow peak} \\
 & & & &  \multicolumn{3}{c}{\hrulefill \ } &  \multicolumn{3}{c}{\hrulefill \ } \\
 & $T^{\rm peak}_{\rm MB}$ & $V_{\rm sys}$ & $\Delta V_{\rm HWHM}$ 
 & $T^{\rm peak}_{\rm MB}$ & $V_{\rm sys}$ & $\Delta V_{\rm HWHM}$
 & $T^{\rm peak}_{\rm MB}$ & $V_{\rm sys}$ & $\Delta V_{\rm HWHM}$ \\ 
Molecule & (mK) & (km~s$^{-1}$) &  (km~s$^{-1}$)  
& (mK) & (km~s$^{-1}$) &  (km~s$^{-1}$) & (mK) & (km~s$^{-1}$) &  (km~s$^{-1}$) \\ \hline

$^{12}$CO (in 2008)\footnotemark[1] & 37$\pm$1 & 47$\pm$1 & 59$\pm$2 
& 23$\pm$2 & 46$\pm$3 & 105$\pm$9 & 26$\pm$3 & 46$\pm$1 & 23$\pm$3 \\ 

$^{12}$CO (in 2010) & 58$\pm$3 & 49$\pm$2 & 47$\pm$3 &
 45$\pm$3 & 49$\pm$2 & 71$\pm$4 & 40$\pm$6 & 46$\pm$1 & 9$\pm$2 \\

$^{12}$CO (in 2011) & 48$\pm$2 & 52$\pm$2 & 42$\pm$3 &
 32$\pm$3 & 56$\pm$3 & 70$\pm$6 & 33$\pm$5 & 48$\pm$1 & 15$\pm$3 \\ 

$^{13}$CO (in 2011)  & 36$\pm$3 & 44$\pm$1 & 36$\pm$3  & ... & ... & ... 
& ... & ... & ... \\ \hline

\end{tabular}

\noindent
\footnotemark[1]Observed on 2008 June 20--21 \citep{ima09}. The spectral synthesis shown in 
that paper is corrected.  
 \end{table*}
 \clearpage

 \begin{figure*}
     \FigureFile(167mm,180mm){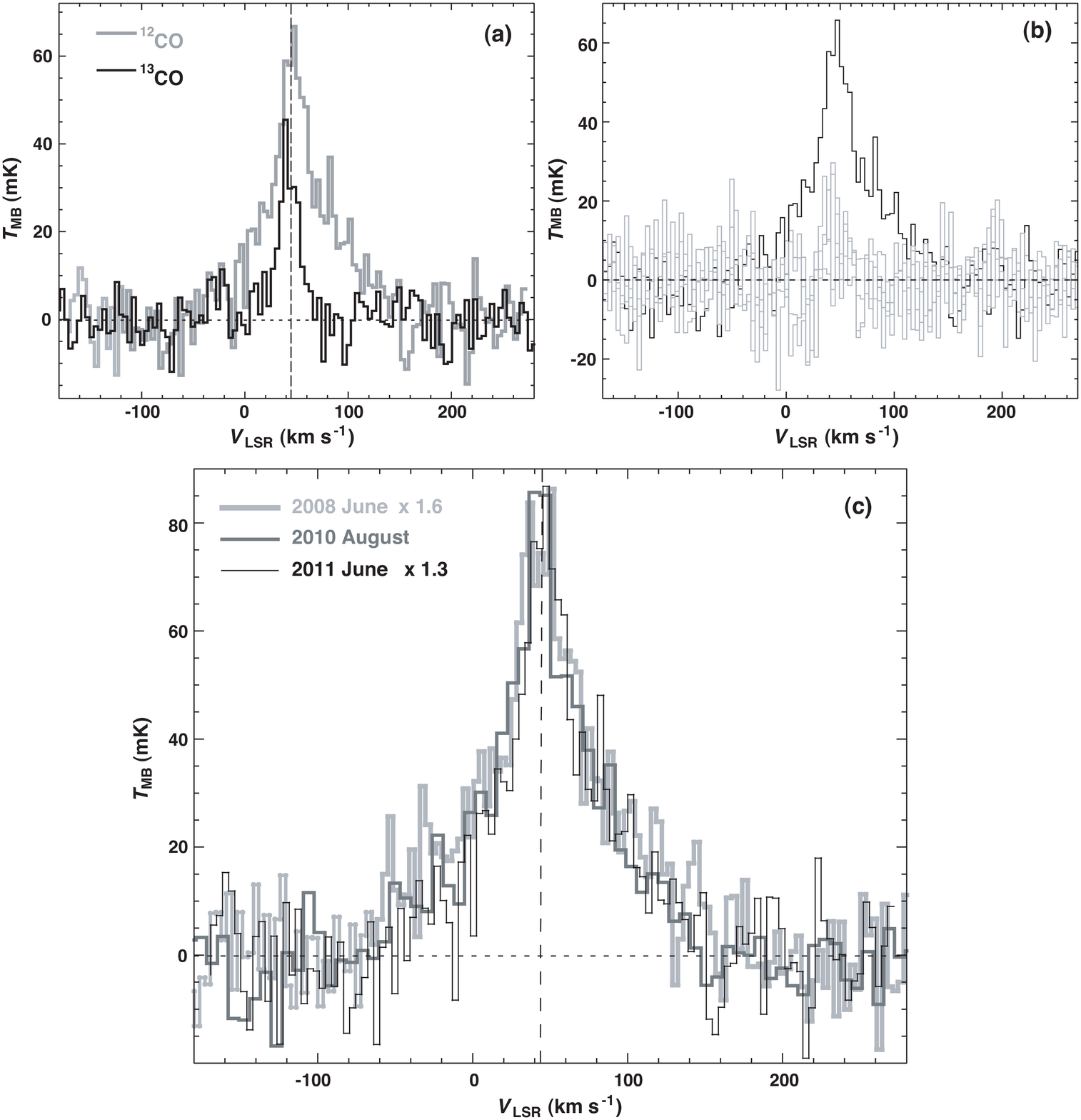}
    \caption{Spectra of the CO \j32 emission lines toward \i1634.  (a) The spectra of the $^{12}$CO 
(a grey thick line) and the $^{13}$CO (a black thin line) emission lines obtained in 2011 June. 
(b) The spectra of $^{12}$CO at the source position (a black line) and the four offset positions 
(gray lines) in the cross-scans conducted in 2011 June. The separation between the on-source 
and offset positions was set to 11\arcsec\ and one of the cross arms was in the direction of the 
major axis of the I16342 jet. (c) The $^{12}$CO spectra obtained in 2008 June (a thick grey line, 
modified from \cite{ima09} and rescaled by a factor of 1.6), in 2010 August (a thick black line), 
and in 2011 June (a thin black line, rescaled by a factor of 1.3). 
The horizontal grey dashed line shows the zero-temperature baseline. }
\label{fig:I16342}
 \end{figure*}
 
\begin{figure}
      \FigureFile(80mm,80mm){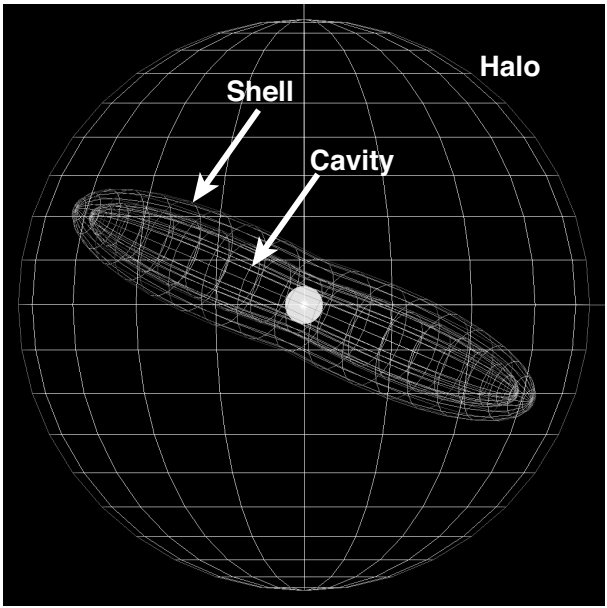}
\caption{{\it Shape} model of \i1634 consisting of a bipolar cavity embedded within a spherical 
halo of CO and dust.  Surrounding the cavity is a thin dense shell which represents material 
swept up by the jet.  The inclination of the cavity/shell is $i=$30\arcdeg.}
\label{fig:SHAPE-geometry}
\end{figure}

\begin{figure*}
      \FigureFile(170mm,120mm){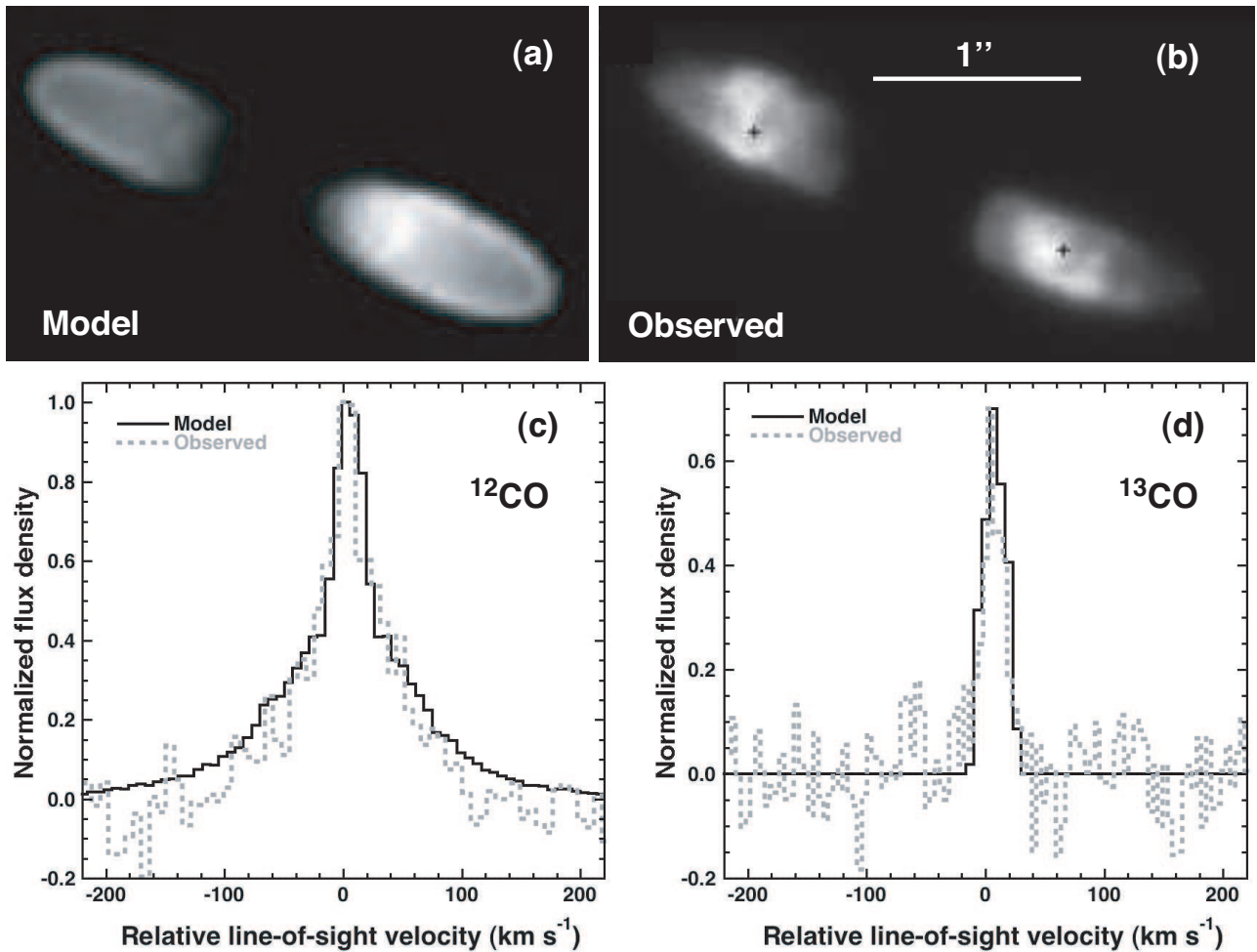}
\caption{Results from the {\it SHAPE} model compared with observations.  
(a) {\it SHAPE} rendering of the optical image model of \i1634.  The image is created using 
radiative transfer within {\it SHAPE}. (b) Near-infrared image of \i1634 cited from \citet{sah05}. 
The image consists of light scattered off dust in the high density shells surrounding a tenuous interior.  
(c)  Comparison between the observed $^{12}$CO spectrum (grey dashed line) and that produced 
from the model (solid black line).  (d) Comparison between the observed $^{13}$CO spectrum (grey dashed) 
and that produced from the model (solid black). }
\label{fig:SHAPE-spectrum}
\end{figure*}
\end{document}